\documentclass[iop]{emulateapj}
\usepackage{graphicx}

\newcommand{\be}{\begin{eqnarray}}
\newcommand{\ee}{\end{eqnarray}}
\newcommand{\lp}{\left(}
\newcommand{\rp}{\right)}

\newcommand{\E}[1]{\times10^{#1}}
\newcommand{\smpy}{ \ M_\odot \ {\rm yr}^{-1}}
\newcommand{\msol}{ \ M_\odot }
\newcommand{\commentOut}[1]{}
\newcommand{\bi}{\begin{itemize}}
\newcommand{\ei}{\end{itemize}}

\shorttitle{ALL DOUBLE WD BINARIES MERGE}
\shortauthors{SHEN}
\slugcomment{Accepted for publication in ApJL}

\begin{document}


\title{Every interacting double white dwarf binary may merge}

\author{Ken J. Shen\altaffilmark{1}}
\altaffiltext{1}{Department of Astronomy and Theoretical Astrophysics Center, University of California, Berkeley, CA 94720, USA.}


\begin{abstract}

Interacting double white dwarf binaries can give rise to a wide variety of astrophysical outcomes ranging from faint thermonuclear and Type Ia supernovae to the formation of neutron stars and stably accreting AM Canum Venaticorum systems.  One key factor affecting the final outcome is whether mass transfer remains dynamically stable or instead diverges, leading to the tidal disruption of the donor and the merger of the binary.  It is typically thought that for low ratios of the donor mass to the accretor mass, mass transfer remains stable, especially if accretion occurs via a disk.  In this Letter, we examine low mass ratio double white dwarf binaries and find that the initial phase of hydrogen-rich mass transfer leads to a classical nova-like outburst on the accretor.  Dynamical friction within the expanding nova shell shrinks the orbit and causes the mass transfer rate to increase dramatically above the accretor's Eddington limit, possibly resulting in a binary merger.  If the binary survives the first hydrogen-rich nova outbursts, dynamical friction within the subsequent helium-powered nova shells pushes the system even more strongly towards merger.  While further calculations are necessary to confirm this outcome for the entire range of binaries previously thought to be dynamically stable, it appears likely that most, if not all, interacting double white dwarf binaries will merge during the course of their evolution.

\end{abstract}



\section{Introduction}

Double white dwarf (WD) binaries are formed via common envelope evolution in binary systems.  The emission of gravitational waves in the closest double WD binaries shrinks their orbits and causes mass transfer between the WDs within a Hubble timescale.  The resulting interaction leads to a broad range of outcomes including explosive transient phenomena and the formation of exotic classes of stars.

Previous work has found that the dynamical stability of mass transfer in these binaries, and thus their ultimate fates, depends on the mass ratio of the donor mass to the accretor mass, $q=M_2/M_1$, and on the timescale for angular momentum feedback from the accretor to the orbit \citep{mns04,krem15a}.  However, systems are generally assumed to be dynamically stable if accretion occurs via a disk, as torques on the extended lever arm of the disk should yield efficient angular momentum feedback.  In this Letter, we challenge this assumption by showing that even disk-accreting double WD binaries with extremely low-mass He WD donors will merge unstably.

The overlying non-degenerate H-rich layers surrounding these low-mass WDs alters the initial phase of mass transfer compared to simplified pure He WD donors \citep{dant06a,kbs12} and leads to classical nova-like outbursts \citep{shen13a}.  Subsequent accretion of the underlying He-rich WD core yields analogous He-powered nova episodes \citep{sb09b}.  Here, we consider the impact of these expanding nova shells on the binary orbit.  We calculate the evolution of two fiducial double WD binaries with masses $0.2+0.5$ and $0.2+1.0 \msol$ and find that these binaries, which were previously assumed to form stably mass transferring AM Canum Venaticorum systems \citep{kili14a}, instead merge due to dynamical friction within the accretor's expanded outer layers.  In Section \ref{sec:energy}, we estimate the shrinking of the orbital separation due to mass ejection via dynamical friction within the common envelope formed by the nova shell.  In Section \ref{sec:mdot}, we provide a prescription for how the donor's mass transfer rate responds to the decrease in orbital separation.  In Section \ref{sec:calcs}, we describe our fiducial calculations, and we consider the implications of our findings in Section \ref{sec:conc}.


\section{Common envelope energetics}
\label{sec:energy}

We begin by estimating the change in the binary separation, $a$, if some or all of the energy to eject the expanding nova shell comes from the orbit via dynamical friction of the two WDs with the common envelope.  This is at best a rough approximation (see \citealt{ivan13a} and references therein); however, we use it here for simplicity and defer analysis with alternative common envelope physics to future work.

The specific energy required to eject material to infinity from the Roche surface of the accreting WD is
\be
	\Phi_{\rm L1} &=&  -\frac{G M_t}{a} \left[ \frac{1/(1+q)}{1 -x_{\rm L1}} + \frac{q/(1+q)}{x_{\rm L1}} - \frac{1}{2}  \lp \frac{1}{1+q} -x_{\rm L1} \rp^2  \right] \nonumber \\
	& \equiv& -\frac{G M_t}{a} f(q) ,
	\label{eqn:pot}
\ee
where the total mass of the system is $M_t = M_1 + M_2$, and $x_{\rm L1}$ is the distance from the center of the donor to the $L_1$ Lagrange point in units of the binary separation.  The function $f(q)$ varies from $1.4-1.9$ for $q=0.1-0.5$.  While this specific energy changes slightly as mass is being ejected from the system, it will remain correct to first order in the ratio of the ejecta mass, $M_{\rm ej}$, to the total mass and thus is sufficient for our purposes of estimation.

Assuming that a fraction $\alpha_{\rm CE}$ of the energy to eject $M_{\rm ej}$ comes from the binding energy of the orbit, we obtain the change in binary separation following the common envelope event,
\be
	\frac{da}{a}  = -\frac{ M_{\rm ej}  }{ M_1 } \left[ \alpha_{\rm CE} \frac{ 2 (1+q) f(q) }{q} + 1 \right] .
\ee
which is again correct to first order in $ \lp M_{\rm ej}/M_t \rp$.
Note that because of the factor of $2/q$, the ratio $da/a$ can be more than an order of magnitude larger than $M_{\rm ej}/M_1$.


\section{Prescriptions for the accretion rate}
\label{sec:mdot}

Dynamical friction during the common envelope event reduces the binary separation, and thus increases the Roche overfill factor, $\Delta = R_2 - R_{\rm L2}$, where $R_2$ is the radius of the donor star, and $R_{\rm L2}$ is the donor's Roche radius, which is approximated by \cite{pacz67} as
\be
	R_{\rm L2} = 0.46 a \lp \frac{ M_2 }{M_t } \rp^{1/3} .
\ee
As a result of the increased overfill factor, the accretion rate rises.  In order to estimate the new perturbed accretion rate, we require a relation for $\dot{M}_2(\Delta)$.  Several such prescriptions exist, including the isothermal approximation \citep{ritt88a}; however, as pointed out by \cite{mns04}, this approximation breaks down when $\Delta$ is larger than the pressure scale height at the donor's photosphere, which will indeed be the case for our post-nova binaries.

We therefore use the adiabatic $\dot{M}$ prescription \citep{pacz72a,savo78a,webb84,mns04}, which assumes an equation of state $P = K \rho^{\Gamma_1}$ with constant $K$ and $\Gamma_1$ along equipotential surfaces.  However, because we are interested in the early phase of mass transfer involving the donor's outer, non-degenerate layers, we cannot assume a constant polytropic index for all donor material as the previously cited studies do.  Instead, we follow the more general expression of \cite{ge10}, which allows for a varying polytropic index within the donor at the complicating expense of requiring a stellar structure integral.  Their approximation to $\dot{M}_2(\Delta)$ is
\be
	\dot{M}_2(\Delta) \sim 2 \pi R_2^2 h(q) \int_{1-\Delta/R_2}^1 \lp \frac{2}{\Gamma_1+1} \rp^{ \frac{\Gamma_1+1}{2 (\Gamma_1-1)}} c_s \rho dx ,
	\label{eqn:mdotdelta}
\ee
where $c_s$ is the sound speed, the factor $h(q)$ is
\be
	h(q) = \frac{q}{ \lp R_{\rm L2}/a \rp^3(1+q)} \frac{1}{ \left[ a_2(a_2-1) \right]^{1/2}} ,
\ee
with $a_2$ given as
\be
	a_2 = \frac{q / \lp 1+q \rp}{  x_{\rm L1}^3} + \frac{1 / \lp 1+q \rp}{  \lp 1-x_{\rm L1} \rp^3} ,
\ee
and we have assumed $\Delta \ll R_2$.  The function $h(q)$ ranges from $1.7-1.4$ for $q=0.1-0.5$.


\section{Fiducial examples}
\label{sec:calcs}

We now utilize the relations in Sections \ref{sec:energy} and \ref{sec:mdot} to calculate the evolution of two fiducial double WD binaries consisting of  an extremely low-mass $0.2 \msol$ He WD donor and $0.5$ and $1.0 \msol$ C/O WD accretors.  The WDs are constructed and evolved with the stellar evolution code MESA\footnote{http://mesa.sourceforge.net, version 7184; default options used unless otherwise noted} \citep{paxt11,paxt13}, beginning with initial metallicities of $0.02$ and main sequence masses of $0.8$ and $7 \msol$ for the $0.5$ and $1.0 \msol$ C/O WD models, respectively.  These WDs are taken from the set of default WD models included with MESA, which employ \cite{reim75a} and \cite{bloe95a} winds to remove mass on the red giant and asymptotic giant branches, leaving hot degenerate cores.  We begin with these cores, turn on chemical diffusion, and allow them to cool for $10^{10}$ yr to approximate the long main sequence timescale of the low-mass companion that will give rise to the $0.2 \msol$ He WD.

We construct our fiducial $0.2 \msol$ WD by halting the evolution of a $1 \msol$ star once it evolves off the main sequence and expands to a radius of $4 \ R_\odot$.  The ``cno\_extras\_o18\_and\_ne22'' nuclear network is utilized, and chemical diffusion is active throughout the simulation.  Mass is removed until only $0.2 \msol$ remains, and the resulting hot WD is allowed to cool for $10^9$ yr.  This yields a He-rich core mass of $0.194 \msol$ and a H-rich shell of $0.006 \msol$.


\subsection{Deriving the evolution of the mass transfer rate}

\begin{figure}
	\plotone{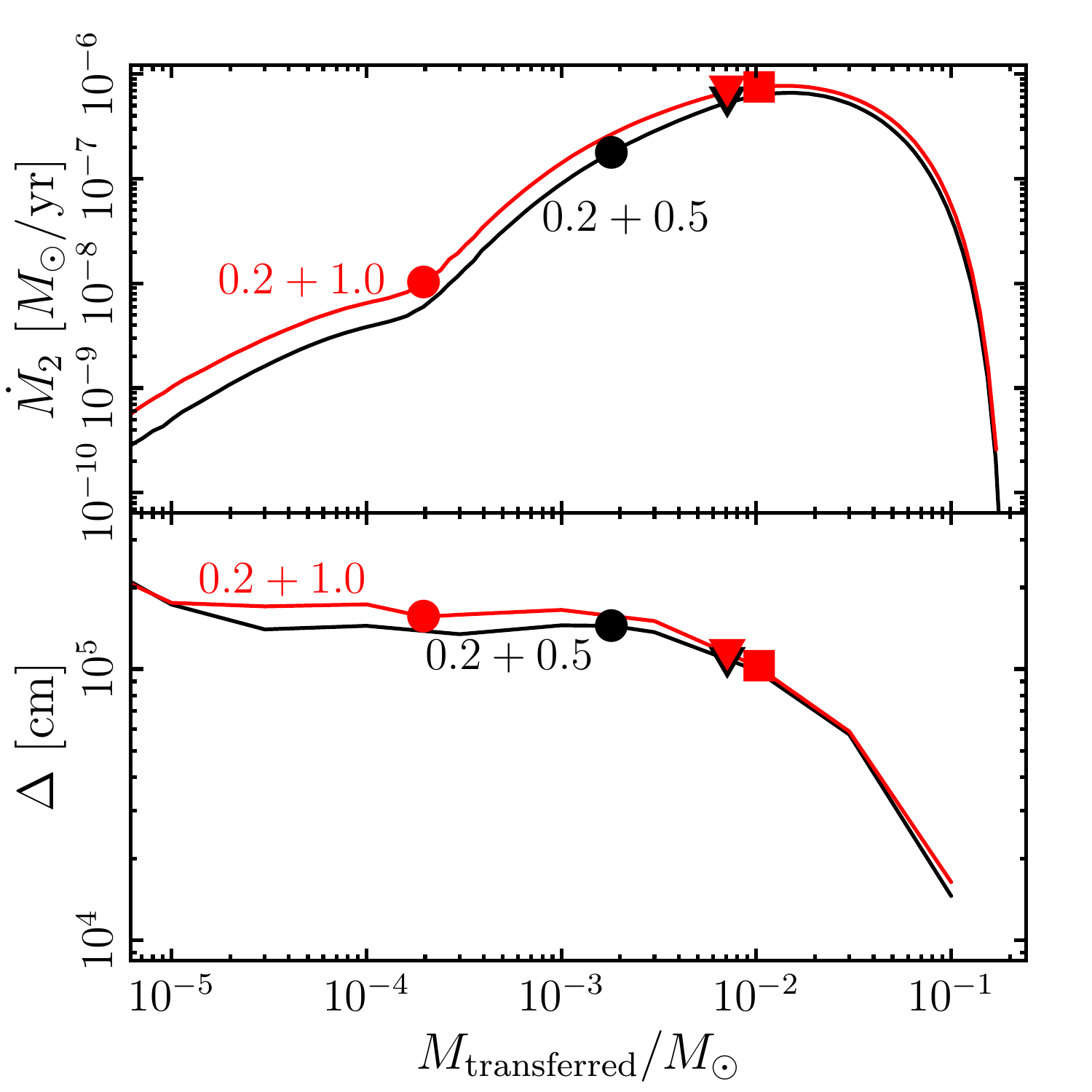}
	\caption{$\dot{M}_2$ (top panel) and $\Delta$ (bottom panel) vs.\ $M_{\rm transferred}$ for our fiducial $0.2+0.5$ and $0.2+1.0 \msol$ binaries, as labeled.  Triangles mark when the H mass fraction of the transferred material drops below 1\%.  These time-dependent calculations do not take into account the binary's response to the nova outburst, which occurs at the bullets in each panel, so the evolution after the nova and subsequent common envelope is merely shown for completeness.}
	\label{fig:mdotanddeltavsm}
\end{figure}

In order to calculate the mass transfer history of these two binaries, we first ignore the accretor's response to accretion and focus only on the donor's response to mass removal.  Since mass transfer is not extremely rapid, it is crucial that we account for the donor's non-adiabatic thermal response.  We first remove mass from the $10^9$-year-old $0.2 \msol$ donor at a constant rate of $10^{-9} \smpy$ for the donor's entire history, then calculate a new mass transfer history based on the donor's radial response.  This is done assuming orbital angular momentum loss is due to gravitational waves and that the angular momentum lost from the donor is transferred efficiently back to the orbit.  This limit is likely applicable to the $0.2+1.0 \msol$ binary because accretion occurs via a disk \citep{ls75} but is only an approximation in the $0.2+0.5 \msol$ system, which undergoes direct impact accretion.  These assumptions yield an accretion rate of
\be
	\frac{\dot{M}_2}{M_2} = -\frac{32G^3}{5c^5} \frac{M_1 M_2 M_t}{a^4} \lp \frac{5}{6} + \frac{\xi}{2} - q  \rp^{-1} ,
\ee
where $\xi \equiv d \ln R_2 / d \ln M_2$.  We then restart the donor evolution with the new time-dependent $\dot{M}_2$, use the donor's response, including the changing value of $\xi$, to calculate a second time-dependent $\dot{M}_2$, and iterate once more.  The maximum difference in $\dot{M}_2$ histories between the penultimate and final iterations is $<1\%$ after the initial $ \sim 10^{-4} \msol$ has been transferred.  The converged $\dot{M}_2$ histories are shown in the top panel of Figure \ref{fig:mdotanddeltavsm} as a function of the transferred mass, $M_{\rm transferred}$, for our two fiducial binaries.  

Also shown in the bottom panel of Figure \ref{fig:mdotanddeltavsm} is the Roche overfill factor corresponding to $\dot{M}_2$ at a given $M_{\rm transferred}$, as defined in equation (\ref{eqn:mdotdelta}).  It ranges from $\simeq 10^5$ cm while the non-degenerate outer layers of the donor are transferred to $\ll 10^5$ cm as the degenerate He core is excavated.


\subsection{The binary's response to the expanding accretor}

The $\dot{M}_2$ history, including the changing accreted composition, is then used to evolve the accreting WDs.  Chemical diffusion is active in these calculations as before.  After $1.8\E{-3} \msol$ has been transferred to the $0.5 \msol$ accretor, nuclear burning triggers convection in the outer $2.8\E{-3} \msol$, and a classical nova-like outburst powered by H-burning begins.  As we will discuss in Section \ref{sec:postCE}, the $0.2+1.0 \msol$ binary will likely remain dynamically stable after its first H-powered outbursts.  We thus instead begin the accretor's evolution for this binary after the initial H-rich $0.007 \msol$ has already been transferred and the accreted material transitions to $>0.99$ He by mass.  Convection then begins on the $1.0 \msol$ accretor after an additional $3.3\E{-3} \msol$ of He-rich material is accreted, resulting in an expanding envelope of $2.9\E{-3} \msol$.  Bullets and squares in Figure \ref{fig:mdotanddeltavsm} show the accretion rates and Roche overflow factors for the first H-powered and He-powered novae, respectively.

As the convective layer heats and expands, it overflows the accretor's Roche lobe and forms a common envelope surrounding the binary.  Even though the majority of the convective layer is expelled, the ejection of even a small fraction of the shell via dynamical friction is sufficient to increase the donor's Roche overfill factor and consequently $\dot{M}_2$ by a tremendous amount.

\begin{figure}
	\plotone{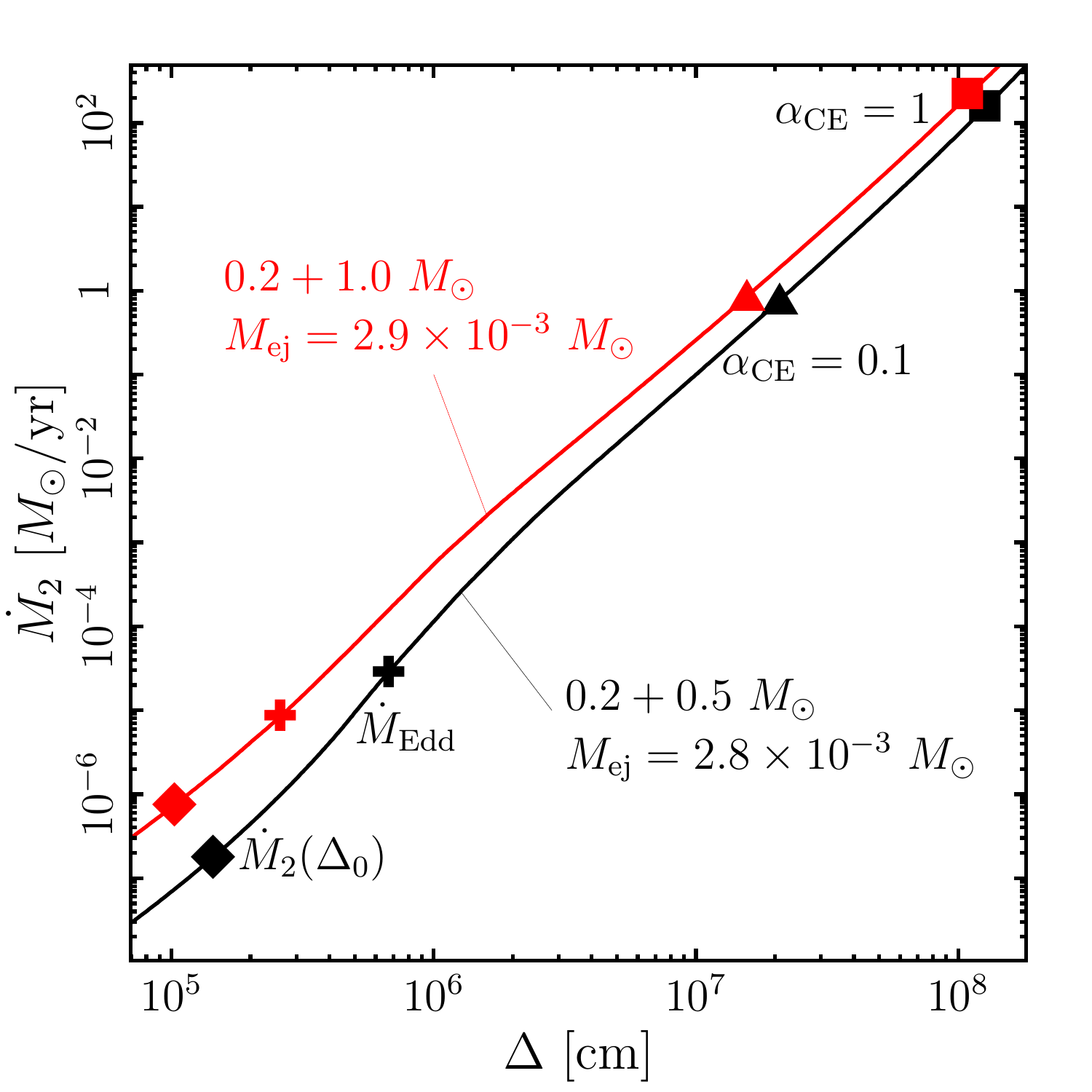}
	\caption{$\dot{M}_2$ vs.\ $\Delta$ for our fiducial binaries, as labeled, at the moment when H-burning and He-burning on the $0.5$ and $1.0 \msol$ accretors, respectively, initiates convection.  Diamonds mark the state of the donor prior to the ensuing common envelope event.  Triangles and squares denote the resulting $\Delta$ and $\dot{M}_2$ if orbital energy is used to eject 10\% and 100\% of the ejecta, respectively.  Crosses represent the Eddington accretion rates for the $0.5$ and $1.0 \msol$ accretors.}
	\label{fig:mdotvsdelta}
\end{figure}

Figure \ref{fig:mdotvsdelta} shows the donor's $\dot{M}_2(\Delta)$ relation at the moment when convection is initiated on the accretor.  Diamonds demarcate $\dot{M}_2(\Delta_0)$ just prior to the common envelope event.  The overfill factors resulting from the decreased orbital separation, assuming $\alpha_{\rm CE}=0.1$ and $1.0$, and the corresponding mass transfer rates are shown as triangles and squares, respectively.  It is clear that the ejection of a relatively small envelope via a common envelope has a drastic effect on these systems, resulting in accretion rates that are highly super-Eddington with respect to the accretors; the Eddington accretion rates, $\dot{M}_{\rm Edd}$, are shown as crosses.


\subsection{The post-common envelope evolution}
\label{sec:postCE}

We now calculate the evolution of our fiducial systems in their post-common envelope, super-Eddington accretion states.  To simplify matters, we approximate the donor's $\dot{M}_2 (\Delta)$ relation as a power-law: $\dot{M}_2 \propto \Delta^b$.  A value of $b=3$ is appropriate for purely degenerate objects, as utilized by \cite{mns04} and others.  Even though the donors in our fiducial binaries are not fully degenerate, the results in Figure \ref{fig:mdotvsdelta} imply that $b \simeq 3$ in these cases as well.  With this simplification, the evolution of $\dot{M}_2$ can be written as
\be
	\frac{d \dot{M}_2/dt}{ \dot{M}_2}  \sim  \frac{ b R_2}{\Delta} \left[ \lp \xi_{\rm rapid} - \frac{1}{3} \rp \frac{ \dot{M}_2 }{M_2} - \frac{ \dot{a} }{a} +  \frac{ 1 }{3} \frac{ \dot{M}_t }{M_t} \right] .
\ee
We have assumed that $\Delta \ll R_2$ and used the \cite{pacz67} Roche lobe relation.  The donor's radial response to the extremely rapid mass loss that follows the common envelope event is $\xi_{\rm rapid}$.

Because the post-common envelope accretion rate is so highly super-Eddington, we further assume that $\dot{M}_t = \dot{M}_2$.  Following a similar energetic argument to that used in Section \ref{sec:energy}, we find
\be
	\frac{\dot{a}}{a}    = \frac{ \dot{M}_2  }{ M_2 }  \left[ 2 \alpha_{\rm CE}  (1+q) f(q)  + 1 \right] ,
\ee
where $f(q)$ is defined in equation (\ref{eqn:pot}).  The evolution of $\dot{M}_2$ now becomes
\be
	\frac{d \dot{M}_2/dt}{ \dot{M}_2}  \sim  \frac{ b R_2}{\Delta} \frac{ \dot{M}_2 }{M_2} \left[ \xi_{\rm rapid}    -   2 \alpha_{\rm CE}  (1+q) f(q)   -    \frac{q+4/3}{1+q} \right] . \nonumber \\
	\label{eqn:dmdotdt}
\ee

The radial response of the donor in the $0.2+0.5 \msol$ binary during the first nova event is initially $\xi_{\rm rapid} \sim 2$ and becomes negative once the electron-degenerate core is exposed.  The term in brackets is thus negative, and the accretion rate becomes larger.  However, for the initial H-rich outbursts of the $0.2+1.0 \msol$ binary, the donor still possesses a relatively large non-degenerate, radiative outer layer, and $\xi_{\rm rapid} \gg 1$.  As a result, the donor is able to rapidly shrink back within its Roche lobe, and the accretion rate decreases back to its equilibrium value.  This system will thus avoid a catastrophic merger during the H-rich nova phase.  However, after further mass transfer exposes the He core, the donor's radial response to the accretor's  subsequent expanding He nova shell becomes negative, and the post-nova accretion rate diverges as in the $0.2+0.5 \msol$ case.  This is the reason we only follow the H-poor evolution for the $0.2+1.0 \msol $ binary in the previous sections: the system may survive the initial H-rich outbursts but will still merge after the first He-rich nova.

With our assumption of a power-law relation for $\dot{M}(\Delta)$, equation (\ref{eqn:dmdotdt}) can be integrated to find that these systems merge on a timescale that is even shorter than the naive post-common envelope accretion timescale, $M_2/\dot{M}_2$, by multiplicative factors of $(\Delta/R_2)$, $(1/b)$, and the inverse of the term in brackets, which ranges from $-2$ to $-6$ depending on the value of $\xi_{\rm rapid}$.  This is roughly the same expression as the timescale for a mildly perturbed double WD binary to return to its equilibrium mass transfer rate \citep{mars05a,stro09a}, although the accretion rate in our fiducial binaries starts at a much higher value and diverges due to the destabilizing effect of dynamical friction.  As a result, the $0.2+0.5$ and $0.2+1.0 \msol$ binaries merge within hours of the common envelope event.


\section{Conclusions and implications}
\label{sec:conc}

\begin{figure*}
	\plotone{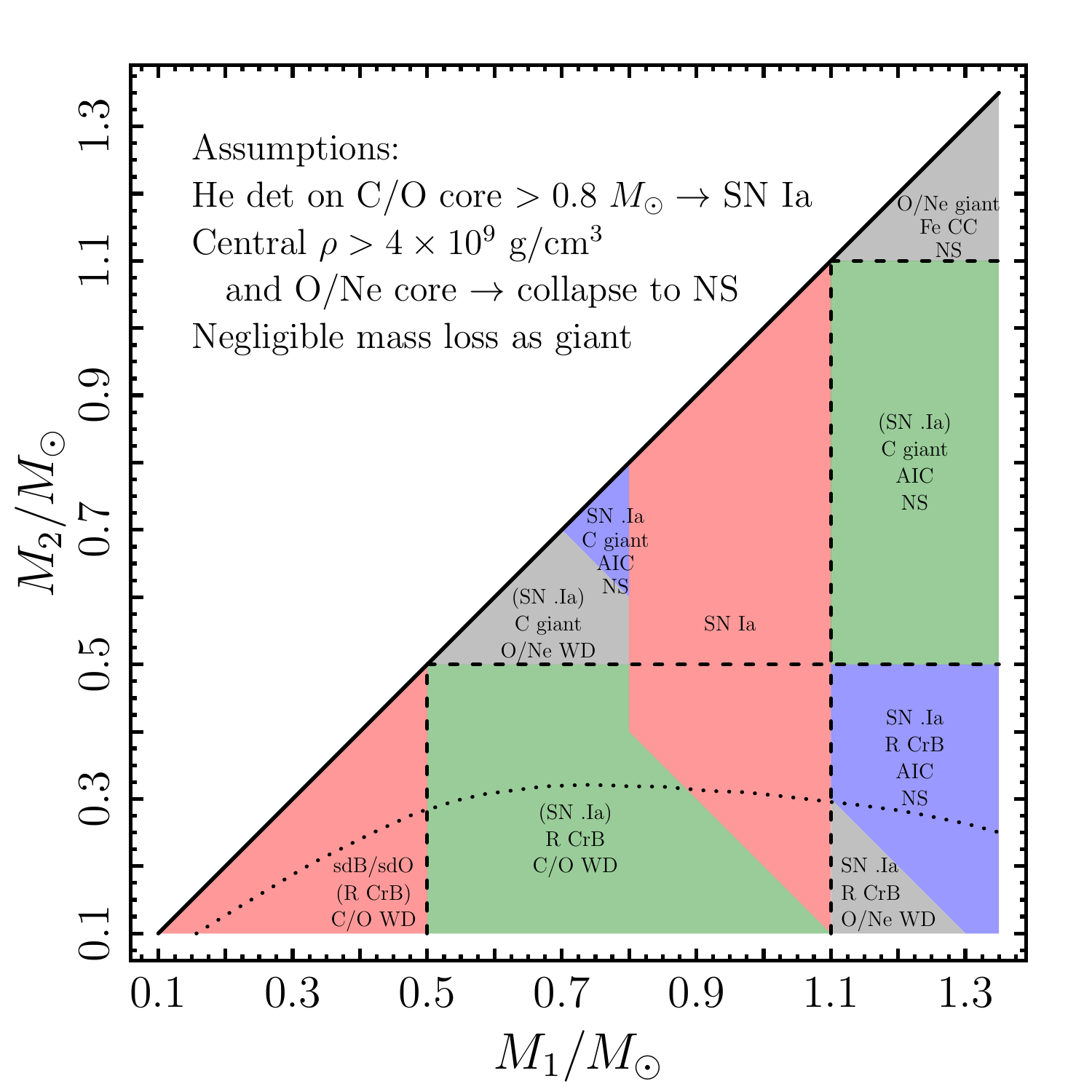}
	\caption{Schematic of interacting double WD binary outcomes.  See Section \ref{sec:conc} for details.}
	\label{fig:mvsm}
\end{figure*}

In this Letter, we have shown that even disk-accreting double WD binaries with extreme mass ratios will merge once common envelope effects during the initial phases of mass transfer are taken into account.  The ejection of the accretor's expanding nova shell causes the binary separation to shrink and the donor to dramatically overfill its Roche lobe, resulting in highly super-Eddington mass transfer rates that lead to a merger.  Further parameter studies covering a range of donor and accretor masses and ages will be required to make definitive conclusions regarding all double WD binaries; however, it appears likely that most, if not all, interacting double WD binaries will merge.

One important consequence of this work concerns the contribution of double WD binaries to the population of AM Canum Venaticorum (AM CVn) systems, which consist of a possibly degenerate He donor and a WD accretor  (see \citealt{solh10a} for a recent review).  One formation channel involves a low-mass He WD donor \citep{pacz67}; however, since we now expect such systems to merge prior to a prolonged phase of He-rich mass transfer, it appears that the AM CVn population must come from a different channel, such as the He-burning star \citep{savo86a,iben87a} or the H star channel \citep{tutu85a,phr03}.  The elimination of the double WD formation channel would help to resolve the large discrepancy between the predicted and observed AM CVn space densities \citep{nele01b,cart13a}, which lends some support to our findings.

A related outcome involves the proposed class of ``.Ia'' supernovae arising from He detonations in stably accreted He shells during the early lives of double WD binaries that were thought to give rise to AM CVns \citep{bild07,sb09b,shen10}.  Such systems now merge before the donor stably transfers a sufficient amount of He to trigger a detonation in a convective He-burning shell.  However, He detonations in double WD binaries may still occur during the violence of double WD mergers if the donor is a He WD or has a large enough surface He layer \citep{guil10,rask12,pakm13a,shen14b}.  If the accretor's core does not subsequently detonate in a double detonation Type Ia supernova (SN Ia; \citealt{livn90,shen14a}), the naked He detonation may still appear as a ``.Ia'' supernova.

The post-nova common envelope events also affect the evolution of higher mass ratio double WD binaries that would have merged even in the absence of dynamical friction within the expanding nova shells.  In \cite{shen13a}, we considered a $0.4 + 1.0 \msol$ binary and examined the effect of the pre-merger nova episodes on the surrounding interstellar medium.  The dense swept-up shells act as absorbing screens for a possible subsequent SN Ia, matching recent observations of blueshifted and/or variable Na absorption features \citep{pata07,ster11}.  However, our calculations did not account for the effect of the common envelopes on the binary evolution.  The ejected envelopes in these higher-mass binaries are smaller than for this Letter's fiducial binaries, so the systems may survive the dynamical friction phases, but the number of ejected shells before the merger may be lower than calculated in \cite{shen13a}.

We conclude by showing a schematic diagram of interacting double WD binary outcomes in Figure \ref{fig:mvsm}, with evolutionary phases in chronological order in each domain.  The long-term evolution of merger remnants that do not explode as SNe Ia during their mergers is assumed to follow the basic framework outlined in \cite{shen12}.  He-rich subdwarf B and O stars are abbreviated as ``sdB/sdO'', R Coronae Borealis stars as ``R CrB'', iron core-collapses as ``Fe CC'', and neutron stars as ``NS''.  Degenerate O/Ne cores that reach high densities are assumed to undergo runaway electron-captures and form a neutron star via accretion-induced collapse (AIC), although the standard picture of accretion from a companion star does not apply here.  SNe Ia are assumed to occur via a core C detonation triggered by a surface He detonation during mergers with $M_1 \gtrsim 0.8 \msol$ \citep{pakm13a,shen14a,shen14b} or via a directly triggered C detonation during ``violent mergers'' of high-mass C/O WDs \citep{pakm10,pakm12b} that do not possess sufficiently massive He surface layers.

Boundaries between regions with differing outcomes are largely qualitative.  Phases in parentheses may not occur for all binaries in the region.  Dashed lines delineate compositional boundaries between He and C/O ($\simeq 0.5 \msol$) and C/O and O/Ne ($\simeq 1.1 \msol$) WDs.  In the binaries below the dotted line, accretion would remain sub-Eddington and they could avoid merging if no post-nova  common envelopes occurred \citep{mns04}.  However, as we have shown in this Letter, it is likely that even these binaries merge due to dynamical friction within common envelopes.


\acknowledgments

We thank Alison Miller and Gijs Nelemans for helpful discussions, Ryan Foley for his persistent encouragement regarding Figure \ref{fig:mvsm}, and the referee for suggestions that helped to clarify this work.  KJS is supported by NASA through the Astrophysics Theory Program (NNX15AB16G).



\end{document}